\documentclass[a4paper,12pt]{article}
\usepackage {epsfig,amssymb,graphics,graphicx}
\usepackage{graphicx}
\usepackage{times}
\usepackage{wrapfig}
\usepackage{natbib}
\begin{document}

\title{Multiphase turbulent interstellar medium:\\
some recent results from radio astronomy}

\author{Nirupam Roy, Max-Planck-Institut f\"{u}r Radioastronomie\\
MPIfR, Auf dem H\"{u}gel 69, Bonn, D-53121, Germany\\
Tel: +49 288 525-491 Fax: +49 228 525-229\\
Email: {\tt nroy@mpifr.de}}

\date{~}
\maketitle

\begin{abstract}
The radio frequency 1.4~GHz transition of the atomic hydrogen is one of the 
important tracers of the diffuse neutral interstellar medium. Radio 
astronomical observations of this transition, using either a single dish 
telescope or an array interferometer, reveal different properties of the 
interstellar medium. Such observations are particularly useful to study the 
multiphase nature and turbulence in the interstellar gas. Observations with 
multiple radio telescopes have recently been used to study these two closely 
related aspects in greater detail. Using various observational techniques, the 
density and the velocity fluctuations in the Galactic interstellar medium was 
found to have a Kolmogorov-like power law power spectra. The observed power law 
scaling of the turbulent velocity dispersion with the length scale can be used 
to derive the true temperature distribution of the medium. Observations from a 
large ongoing atomic hydrogen absorption line survey have also been used to 
study the distribution of gas at different temperature. The thermal steady 
state model predicts that the multiphase neutral gas will exist in cold and 
warm phase with temperature below $200$ K and above $5000$ K respectively. 
However, these observations clearly show the presence of a large fraction of 
gas in the intermediate unstable phase. These results raise serious doubt about 
the validity of the standard model, and highlight the necessity of alternative 
theoretical models. Interestingly, numerical simulations suggest that some of 
the observational results can be explained consistently by including the 
effects of turbulence in the models of the multiphase medium. This review 
article presents a brief outline of some of the basic ideas of radio 
astronomical observations and data analysis, summarizes the results from recent 
observations, and discusses possible implications of the results. 
\end{abstract}

\noindent{\bf Key words:} ISM: atoms -- ISM: general -- ISM: structures -- radio lines: ISM -- turbulence

\newpage


\noindent{\bf \Large Introduction}\vspace{0.25cm}

\noindent The interstellar space in a galaxy is far from empty. Even if the 
density is very low ($\sim 10^{-4} - 10^6$ cm$^{-3}$), the region between the 
stars is filled with the interstellar medium (ISM) consisting of gas, dust, 
charged particles and magnetic fields. Astronomers inferred the existence of 
interstellar gas after the discovery of stationary absorption lines of ionized 
calcium (Ca~{\sc ii}) towards the spectroscopic binary $\delta$~Orionis 
\citep{hart1904}. Any spectral line from stars of a spectroscopic binary 
system should have a periodic wavelength shift due to the Doppler effect for 
the motion of the stars. The stationary Ca~{\sc ii} absorption line was hence 
taken as an indication of interstellar gas. Edward Emerson Barnard first 
carried out a systematic study of the Galactic ISM, and proposed the existence 
of ``intervening opaque masses'' between the stars to explain ``dark nebula''. 
He used his expertise of astrophotography to produce the first images of dark 
nebulae and published the first catalogue of such dark clouds \citep{barn19}. 
Over time, further imaging and spectroscopic observations, made the existence 
of widespread ISM, including gas and dust clouds, quite evident.

It is now well established that the process of star formation through the 
collapse of protostellar clouds is never totally efficient. This leads to the 
existence of residual gas around the stars. The radiation and mechanical 
energy inputs from the stars, in turn, influence the properties of the ISM. 
Processes like stellar winds, episodic ejection of mass during the stellar 
evolution and supernova explosions transfer material and energy from the stars 
to the ISM. Similarly, ultraviolet radiation from hot, young stars and cosmic 
rays heat and ionize the ISM. Due to all these processes, the ISM is strongly 
coupled to the stars in a galaxy, and play a crucial role between the stellar 
and the galactic scales. Hence, understanding the properties of the ISM is 
very important from many considerations in astrophysics.

Multiphase nature and turbulence are two of the key ingredients of the ISM 
physics. In the standard model, multiple phases of the ISM, with different 
densities, temperatures and ionization states, coexist in rough thermal 
pressure equilibrium \citep{fiel65,fiel69}. In a multiphase medium, 
the cold dense neutral gas (cold neutral medium; CNM) is embedded in either 
neutral or ionized low density warm gas (warm neutral and ionized medium; 
WNM/WIM), which again may be embedded in much lower density hot ionized medium 
(HIM). Considering different cooling and heating processes, it can be shown 
that the neutral atomic hydrogen gas (H~{\sc i}) can be in thermal steady state 
in one of the two stable ranges of temperature: (i) $\sim 40 - 200$ K for CNM 
and (ii) $\sim 5000 - 8000$ K for WNM. The H~{\sc i} at intermediate 
temperatures is unstable, and, due to runaway heating or cooling, moves to 
either cold or warm phase \citep{mcke77,wolf95,wolf03}. The ISM (of the Milky 
Way as well as of other galaxies) is also known to have a clumpy, self-similar, 
hierarchical structure over several orders of magnitude in scale 
\citep{lars81,falg92,heit98}. Various direct and indirect observations suggest 
that these structure extends down to $\lesssim 10$ AU scales 
\citep{hill05,dutt14}. On the theoretical front, the ISM is known to be 
turbulent, and hence is expected to exhibit such density structures and 
velocity fluctuations over a wide range of scales \citep{crov83,gree93,laza00}. 
However, the nature of the turbulence in different phases of the ISM is not yet 
understood properly.

This article presents an overview of some recent results, from radio 
astronomical observations, which have important implications for our current 
understanding of the multiphase turbulent ISM. A significant fraction of the 
observations, on which these results are based, were carried out with the Giant 
Metrewave Radio Telescope \citep[GMRT;][]{swar91}, a world-class radio 
telescope located near Pune, Maharashtra, in India. A few important aspects of 
the observational techniques in radio astronomy, and a brief overview of the 
GMRT are outlined in the ``Observation and Analysis'' section. The results, 
related to the ISM turbulence and the multiphase nature, are summarized in 
sections titled ``Turbulence in the diffuse neutral ISM'' and ``Temperature of 
the diffuse neutral ISM'' respectively. Finally, a discussion on the broader 
aspects of these results, and on possible future directions are presented in 
the ``Discussions and Conclusions'' section.\\


\noindent{\bf \Large Observation and Analysis}\vspace{0.25cm}

\begin{figure}[t]
\begin{center}
\includegraphics[width=3.5cm, height=15cm, angle=-90.0]{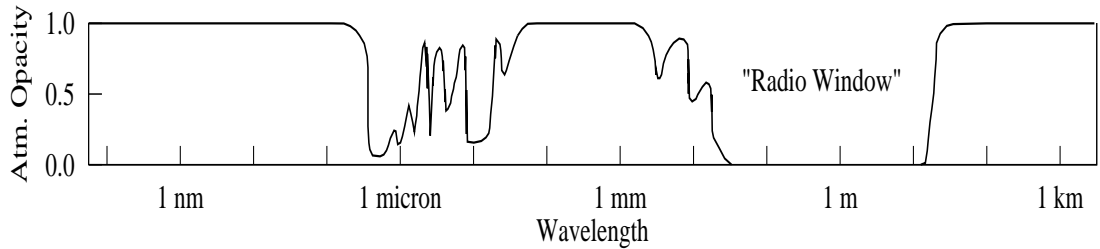}
\caption{\label{fig:atop} 
\small{A toy model of atmospheric opacity to electromagnetic radiation at different wavelength. Note the low opacity for optical ($\lambda \sim 400 - 700$ nm) and radio waves (few cm to $\sim 10$ m).}}
\end{center}
\vspace{-0.5cm}
\end{figure}

\noindent Historically, astronomy was confined to observations of the sky 
mostly with optical telescopes, probing only a narrow window of visible lights 
of the whole electromagnetic spectrum. With technological progress and 
developments, today astronomers are, however, capable of observing, using 
ground- and space-based telescopes, across the whole spectrum -- $\gamma$-rays, 
X-rays, ultraviolet, visible, infrared, microwave and radio emission 
originating in different astrophysical processes from a variety of sources. Due 
to low or negligible atmospheric opacity at the radio ``window'' ($\lambda 
\approx$ a few cm to about 10~m; see Figure~\ref{fig:atop}), observations at 
these wavelengths are possible to carry out from ground-based telescopes.

Radio astronomy emerged as a subject in the first half of the last century, 
with the pioneering work of Karl Guthe Jansky, who first reported the detection 
of celestial radio signal \citep{jans33}. His contribution has been recognised 
in many ways - from the unit of flux density Jansky ($1$ Jy $=10^{-26}$ 
W~m$^{-2}$~Hz$^{-1}$), to recent renaming of the upgraded radio telescope as 
the Karl G.~Jansky Very Large Array (VLA). Over time, radio observations have 
revealed thermal and non-thermal emission from known as well as new classes of 
sources including the Sun, supernova remnants, pulsars, regions of ionized 
hydrogen (H~{\sc ii} regions) associated with star formation, active galactic 
nuclei, radio galaxies and clusters of galaxies. To get an overview of the 
exciting results from radio astronomical studies, see \citet{burk14}. For the 
purpose of this article, let us now consider a very important radio frequency 
transition, the famous H~{\sc i} 21 cm line, in greater details.\\


\noindent{\it \large Atomic hydrogen 21 cm radiation}\vspace{0.25cm}

\begin{wrapfigure}{r}{0.51\textwidth}
\vspace{-0.8cm}
\begin{center}
\includegraphics[angle=-90, width=0.5\textwidth]{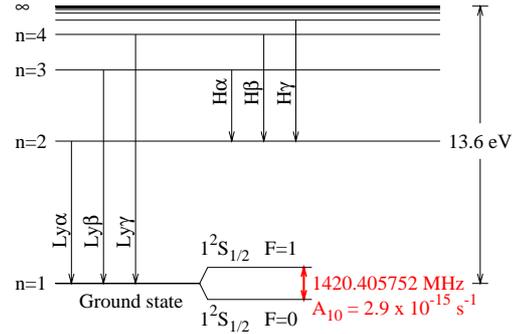}
\end{center}
\vspace{-0.3cm}
\caption{\label{fig:hien} 
\small{A simplified representation of the energy levels of atomic hydrogen showing the hyperfine splitting of the ground state. Ly-$\alpha$ and other higher energy level transitions may play a role in bringing equilibrium between the hyperfine states.}}
\end{wrapfigure}

\noindent In the radio frequencies, the most useful tracer of the ISM is the 
H~{\sc i} 1.4~GHz (21 cm) line. This is not only because hydrogen is the most 
abundant element in the ISM, but also because it is possible to extract a lot 
of useful information (e.g. temperature, column density, velocity, magnetic 
field strength) about the physical properties of the ISM from H~{\sc i} 21 cm 
observations. In 1944, Hendrik van de Hulst predicted the possibility of 
detecting H~{\sc i} 21 cm radiation of celestial origin \citep{huls45}, and it 
was first observed by \citet{ewen51}. This line emission (or absorption) is 
caused by the transition between the two hyperfine states of the 
$1^2S_{\frac{1}{2}}$ ground state of hydrogen. These two states with parallel 
and anti-parallel electron/proton spin configuration (i.e. total spin angular 
momentum $F=1$ and $0$) have slightly different energy (see 
Figure~\ref{fig:hien}). The energy difference $h\nu$ corresponds to the 
frequency of the transition $\nu_{10} = 1420.405752$ MHz ($\sim 21.1$ cm).

The 21 cm transition of neutral hydrogen has an extremely small probability of 
$2.9\times10^{-15} {\rm s}^{-1}$. So, the natural lifetime of the excited 
state for a single isolated atom is about $10$ million years. But, because of 
the processes like collisions and interaction with the background radiation 
field, the lifetime can be considerably shortened. This, and the fact that the 
total number of atoms in the ISM is quite large, make the 21 cm emission (or 
absorption against a background source) readily detectable.

In thermodynamic equilibrium, the number density of atoms in the upper and 
lower energy state are related by the Boltzmann distribution, $n_1 = 3n_0 
\exp(-h\nu_{10}/kT)$. Even if the gas is not in thermodynamic equilibrium, one 
can define a characteristic temperature, called the spin temperature ($T_s$), 
that satisfies the observed distribution between the two spin states.

Either collision or radiative mechanisms may cause direct transition between 
the two hyperfine states. Alternatively, higher energy levels of hydrogen atom 
can be excited from one of the hyperfine states, followed by de-excitation to 
the other hyperfine state via Ly-$\alpha$ or higher Lyman lines transitions 
(see Figure~\ref{fig:hien}). The probability of transition via Ly-$\alpha$ 
photons depends on the intensity profile of radiation near the Ly-$\alpha$ 
frequency. Radiation which has been scattered many times in the ISM will have 
a profile dependent on the kinetic temperature ($T_k$) of the gas. This may 
tend to couple $T_s$ to $T_k$, particularly in low density gas. In high density 
regions, the collisional mechanism dominates, and may similarly couple $T_s$ 
to $T_k$ \citep{fiel58}. Whether or not $T_s$ is same as $T_k$, for a constant 
$T_s$ along the line of sight, it can be shown \citep{kulk88} that 
$T_B(\nu) = T_s[1-e^{-\tau(\nu)}]$, where $T_B(\nu) = I(\nu)c^2/2k\nu^2$ is 
the observed brightness temperature, $\tau(\nu)$ is the optical depth at 
frequency $\nu$, $I(\nu)$ is the intensity, $k$ is the Boltzmann constant and 
$c$ is the speed of light. At low optical depth limit, the spin temperature 
will be $T_s(\nu) \approx T_B(\nu)/\tau(\nu)$.\\


\noindent{\it \large Radio telescopes}\vspace{0.25cm}

\begin{figure}[t]
\begin{center}
\includegraphics[width=15cm,angle=0.0]{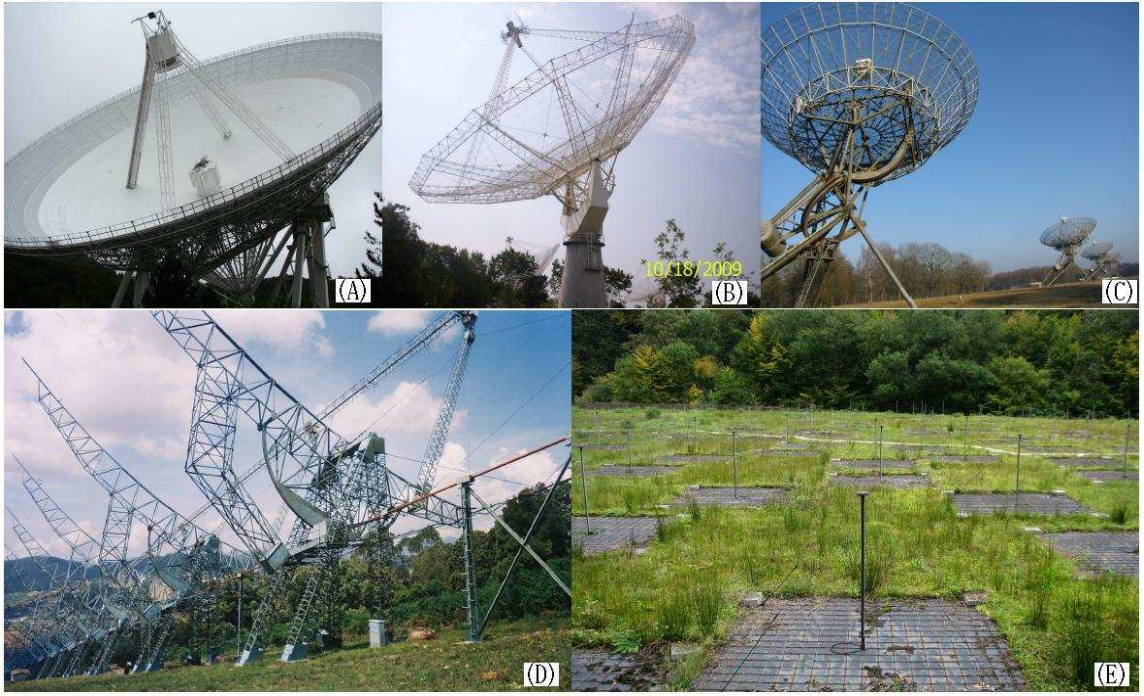}
\caption{\label{fig:scop} 
\small{(A): 100 m diametre single dish Effelsberg Radio Telescope in Germany; (B): One of the 30 dishes of the GMRT array; (C): A few of the 25 m dishes of the Westerbork Synthesis Radio Telescope (WSRT; a 14 element 2.7 km East-West array) in the Netherlands. Not all radio telescopes are dish shaped. (D): The Ooty Radio Telescope (ORT) is a $530 {\rm ~m}\times30 {\rm ~m}$ cylindrical paraboloid on equatorial mount; (E): Part of a Low Frequency Array (LOFAR) station in Germany consists of the low band dipole antennas. See text in the ``Radio Telescope'' sub-section for details.}}
\end{center}
\vspace{-0.5cm}
\end{figure}

\begin{figure}[t]
\begin{center}
\includegraphics[width=15cm,height=5cm,angle=0.0]{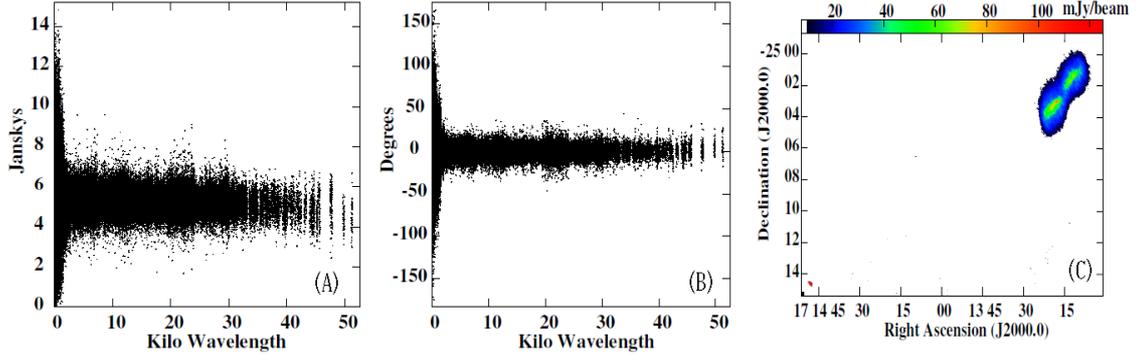}
\caption{\label{fig:data} 
\small{Calibrated visibility data and image from the GMRT 610 MHz observation of the field containing the bright unresolved source NVSSJ$171447-251435$. (A): Visibility amplitudes and (B): phases as a function of baseline separation; (C): Image of a part of the field, created by inverse Fourier transform of the data, shows a double-lobed radio galaxy at the top right corner, and the point source with $\sim 5$ Jy flux density at the bottom left corner. The flux density scale of the image is saturated at $\sim 125$ mJy to properly show the faint extended radio galaxy.}}
\end{center}
\vspace{-0.5cm}
\end{figure}

\noindent Though there may be considerable varieties (see, e.g., LOFAR and ORT 
in Figure \ref{fig:scop}), radio telescopes are generally parabolic antennas to 
receive and detect radio frequency electromagnetic waves. A radio telescope may 
consists of either a single element (``single dish'' telescope), or multiple 
elements (``array'' telescope). The technique of interferometry is routinely 
used in radio astronomy to achieve better sensitivity and angular resolution 
(or the ``beam'') by combining signals from widely separated elements of an 
array telescope. The resolution of a single dish telescope is $\approx 
\lambda/D$, while the same for an interferometer is $\approx \lambda/B$ for an 
observing wavelength of $\lambda$, dish diametre $D$ and maximum separation 
between antennas $B$ (known as interferometer ``baseline''). A single dish 
telescope is sensitive to the total intensity $I(l,m)$ towards the direction of 
observation $(l,m)$ on the sky. An interferometer, on the other hand, directly 
measure the Fourier transform of $I(l,m)$, 
\begin{equation}
V(u,v) = \int\int I(l, m)e^{-2\pi i(ul+vm)}dldm\,.
\end{equation}
Each combination of antenna pair sample the real and the imaginary part (or 
amplitude and phase) of the visibility function $V(u,v)$ at an inverse angular 
separation $(u,v)$ corresponding to projected baseline separation perpendicular 
to the source direction. As an example, Figure~\ref{fig:data} (A) and (B) shows 
the calibrated visibility data for a GMRT 610 MHz observation. At a given time, 
an array with $N$ elements sample the visibility function on $N(N-1)/2$ points 
on the $(u,v)$ plane. During the course of the observation, as the apparent 
position of the source on the sky changes with time, the projected baseline 
separation also changes. This technique of sampling the $(u,v)$ plane is known 
as the aperture synthesis. The maximum baseline for such arrays may be from few 
km to tens of km, and even few thousand km for very long baseline 
interferometry (VLBI). The longest baseline of the Very Long Baseline Array 
(VLBA) of the National Radio Astronomy Observatory (NRAO), for example, is 
$\sim 8600$ km (Mauna Kea, Hawaii to St. Croix, Virgin Islands). Inverse 
Fourier transform of $V(u,v)$ is used to make the ``radio image'' of the sky. 
Figure~\ref{fig:data} (C) also shows the corresponding image created using the 
Astronomical Image Processing Software (AIPS). Details of methods for either 
single dish or interferometric observations, and standard data analysis 
procedures are beyond the scope of this article. One may see \citet{thom01}, 
\citet{wils13}, and references therein, for a comprehensive discussion on 
these topics.\\


\noindent{\it \large The Giant Metrewave Radio Telescope}\vspace{0.25cm}

\noindent The GMRT is, at present, world's largest low frequency radio 
telescope. It is located near the village Khodad in Maharashtra, approximately 
80 km from Pune. The GMRT consists of 30 parabolic dishes, each with a diametre 
of 45 m. The telescope operates at $\sim 150$, $235$, $325$, $610$ and $1420$ 
MHz. Half of the antennas are placed within central $1 {\rm ~km}\times1 {\rm 
~km}$ area, and the rest are placed in roughly a ``Y'' shaped array (with 
largest baseline separation of $\sim 30$ km) for optimum sampling of the 
$(u,v)$ Fourier domain. The unique feature of GMRT, apart from its high angular 
resolution and low frequency capabilities, is the simultaneous sensitivity of 
the array to both large and small scale structures due to this hybrid array 
design. The parabolic surface of the antennas consist of wire mesh which are 
good reflector at the operating frequency range of the GMRT. All the antennas 
can be moved to point at different directions in the sky using an 
altitude-azimuth mounting system. The signals from all the antennas are 
transmitted to the array control building using optical fiber network, and 
combined in the correlator system to record the visibilities in a digital 
format. The angular resolution of the GMRT at 1.4~GHz is $\lesssim 2$ 
arcsecond. It is possible to easily achieve a spectral resolution of $< 2$ kHz 
per channel ($\sim 0.4$ km~s$^{-1}$). More details on GMRT, including all 
technical details are summarized by \citet{swar91}.\\


\newpage

\noindent{\it \large Emission and absorption spectra}\vspace{0.25cm}

\noindent A single dish radio telescope with receiver covering the 21 cm line 
(``L band'') can be used to observe radio emission from the Galactic H~{\sc i}. 
Such observations typically have coarse angular resolution (tens of arcminute 
beam size). One may also use a single dish telescope to observe H~{\sc i} in 
absorption towards background continuum sources. However, to do that, one must 
carefully model the emission spectra from surrounding pointing, and subtract 
the contamination from emission within the beam to estimate the absorption 
spectra. Thus, effectively one assumes that the H~{\sc i} emission is smooth 
over the scale of a few beam size. This assumption may not necessarily be 
correct. A much reliable way to get the absorption spectra is to carry out 
interferometric observations. With the typical resolution of few arcsecond, 
the diffuse H~{\sc i} emission is resolved out, and one gets uncontaminated 
absorption spectra towards background continuum sources. Following standard 
data reduction steps (removing bad data and radio frequency interference, 
calibrating instrumental and atmospheric effects, imaging of the visibility 
data using inverse Fourier transform, converting absorption intensity to 
optical depth $\tau$ etc.) one finally gets both the H~{\sc i} emission and 
absorption spectra $T_B(\nu)$ and $\tau(\nu)$, from single dish and/or 
interferometric observations, with certain spectral resolution $\Delta \nu$. 
These spectra are also often expressed as a function of physically more 
meaningful Doppler velocity ($v$), instead of observed frequency ($\nu$).\\


\noindent{\it \large Derived physical quantities}\vspace{0.25cm}

\noindent For an isothermal H~{\sc i} cloud, both the emission and the 
absorption spectra can be modelled as Gaussian profile along the velocity axis. 
The central velocity of the Gaussian represents the line of sight component of 
the ``bulk'' velocity, and is closely related to the Galactic rotation. The 
velocity width of the line has contribution from thermal and non-thermal 
velocity dispersion. In absence of any non-thermal motion, the line width 
provides $T_k = (m_H/k) (\sigma/1{\rm ~km~s}^{-1})^2$ K. In reality, however, 
due to non-thermal broadening, it only provides an upper limit 
($T_{k,\rm max}$) of the kinetic temperature. The H~{\sc i} column density 
(i.e. the integrated number density along the line of sight) is given by,
\begin{eqnarray}
N({\rm H~I}) = 1.823\times10^{18}T_s\int{\tau(v)dv} \approx 1.823\times10^{18}\int{T_B(v)dv} 
\end{eqnarray}
where $N({\rm H~I})$ is in cm$^{-2}$ and $v$ is in km s$^{-1}$ 
\citep{kulk88,dick90}. The spin temperature, $T_s$ is given by,
\begin{equation}
T_s = \frac{N({\rm H~I})}{1.823\times10^{18}\int{\tau(v)dv}} \approx \frac{\int{T_B(v)dv}}{\int{\tau(v)dv}}
\label{eqn:tsdef}
\end{equation}
where $\tau(\nu)$ is generally determined from H~{\sc i} absorption spectrum 
towards a background source, and $N({\rm H~I})$ is determined from the 
emission spectrum for a nearby line of sight. 

In practice, if the gas is not homogeneous (i.e. having variation of density 
and temperature, along and transverse to the line of sight), or the optical 
depths are large ($\tau \gtrsim 1$), Equation~\ref{eqn:tsdef} can not be used 
to determine `the' spin temperature easily. For example, multiple optically 
thin ($\tau \lesssim 1$) components along the line of sight will only give a 
column-density-weighted harmonic mean temperature of the individual components. 
In general, there may not be any unique interpretation of the data, although 
modelling the spectra with many Gaussian components (corresponding to multiple 
``clouds'') is common in such situations \citep{mebo82,ht03a,ht03b}. 
For extreme situations with very complicated profiles, where multi-Gaussian fit 
may be highly non-trivial and degenerate, one may still extract some useful 
information by computing $T_s(v) = T_B(v)/\tau(v)$ for each spectral channel. 
Note that the analysis method outlined above can be extended easily for 
imaging widespread emission or absorption against extended background sources. 
In those cases, $\tau(\nu)$ and $T_B(\nu)$ will also be functions of sky 
coordinates $(l,m)$.

\begin{wrapfigure}{r}{0.56\textwidth}
\vspace{-0.8cm}
\begin{center}
\includegraphics[angle=-90, width=0.55\textwidth]{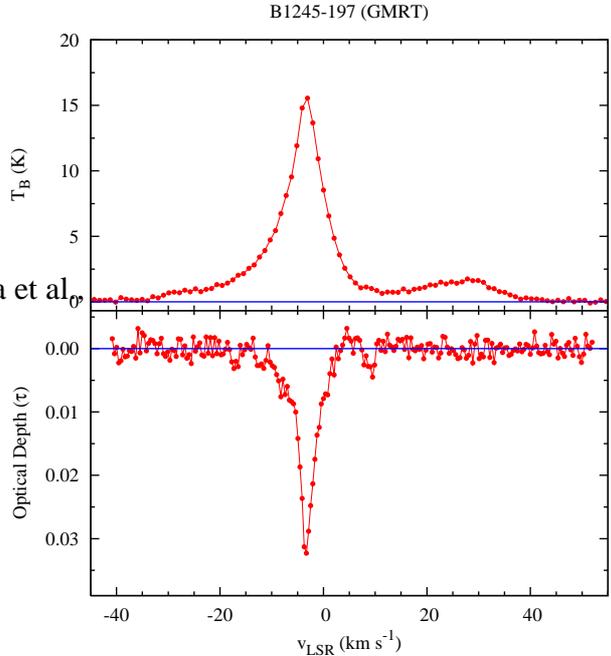}
\end{center}
\vspace{-0.3cm}
\caption{\label{fig:spec} 
\small{Profiles of the observed H~{\sc i} 21 cm emission and absorption towards $l= 302^\circ$, $b=+43^\circ.5$. The emission spectrum is taken from the LAB survey, and the absorption spectrum is obtained with the GMRT (also see the section titled ``Temperature of the diffuse neutral ISM'').}}
\end{wrapfigure} 

Figure (\ref{fig:spec}) shows an example of H~{\sc i} absorption and emission 
spectra towards quasar PKS~$1245-19$, a radio bright, unresolved extragalactic 
source ($l= 302^\circ$, $b=+43^\circ.5$ in the Galactic coordinate system). The 
emission spectrum is taken from the Leiden/Argentine/Bonn Survey 
\citep[LAB;][]{hart97,baja05,kalb05}, and the absorption spectrum is from the 
GMRT. One can clearly see multiple emission and absorption components at 
different velocity ($v_{\rm LSR}$) with respect to the local standard-of-rest 
(LSR) frame. The H~{\sc i} column density, $N({\rm H~I})$, computed from the 
integral of the emission profile, is $\sim 3.7 \times 10^{20}$ cm$^{-2}$, and 
the average spin temperature is $\langle T_s \rangle = 1288 \pm 41$ K. Both the 
emission and absorption spectra can also be modelled as a set of multiple 
Gaussian components. There are some broad emission components without 
corresponding absorption components. Such components are believed arise from 
higher temperature gas (i.e. WNM), with optical depth lower than the detection 
limit of the observation \citep{radh72,ht03a,ht03b}. For this particular line 
of sight, from $\langle T_s \rangle$, and $T_{k,\rm max}$ values derived from 
the line widths \citep{nroy13b}, one can conclude that only $\lesssim 16\%$ of 
H~{\sc i} is in the CNM phase (with $T_s \lesssim 200$ K).\\


\noindent{\bf \Large Turbulence in the diffuse neutral ISM}\vspace{0.25cm}

\noindent Indications of turbulence in the diffuse neutral ISM come mainly from 
observations of small scale fluctuations of H~{\sc i} emission as well as of 
optical depth. Recently, it has been clearly shown from an ongoing H~{\sc i} 
absorption survey \citep[][also see the section titled ``Temperature of the 
diffuse neutral ISM'']{nroy13a,nroy13b} that, for the diffuse H~{\sc i}, the 
non-thermal broadening contributes significantly (i.e. comparable to the 
thermal broadening) to the observed line width. There are direct observational 
evidences of AU scale (milliarcsec angular scale) structures in H~{\sc i} 
opacity fluctuations from VLBI observations \citep{brog05,nroy12,dutt14}, 
and also from variation of absorption towards high proper motion pulsars 
detected using multi-epoch observations \citep{frai94}. \citet{desh00} 
reported scale-free opacity fluctuations at pc and sub-pc scales using 
interferometric observations towards extended sources. Similarly, H~{\sc i} 
emission studies have revealed scale-free structures of the neutral ISM in the 
Milky Way and other nearby galaxies \citep{crov83,gree93,west99,dutt09}. Most 
of these studies use second order statistics (e.g. structure function, power 
spectrum) of intensity and opacity fluctuations, and compare those with 
analytical or numerical models \citep{kolm41,iros64,krai65,gold95} to constrain 
the nature of the turbulence.

Power spectra of intensity or opacity fluctuations are related directly to the 
underlying density fluctuations, but also got affected by the velocity 
fluctuations \citep{laza00}. The power spectrum, which is the Fourier 
transform of the autocorrelation function, is given by
\begin{equation}
P(u,v)=\int\int\xi(l, m)e^{-2\pi i(ul+vm)}dldm
\end{equation}
where $\xi(l,m)$ is the autocorrelation function of the quantity of interest 
(intensity, opacity, density or velocity fluctuations). The function $\xi(l,m)$ 
is also closely related to the structure function $S(l,m)$. For the intensity 
fluctuation, $\xi$ and $S$ can be written as
\begin{eqnarray}
\xi(l-l^{\prime}, m-m^{\prime})&=&\langle\delta I(l,m)\delta I(l^{\prime}, 
m^{\prime})\rangle \nonumber \\
S(l-l^{\prime}, m-m^{\prime})&=&\langle[I(l,m) - I(l^{\prime}, m^{\prime})]^2\rangle \, .
\end{eqnarray}
\begin{wrapfigure}{r}{0.56\textwidth}
\vspace{-0.8cm}
\begin{center}
\includegraphics[angle=-90, width=0.55\textwidth]{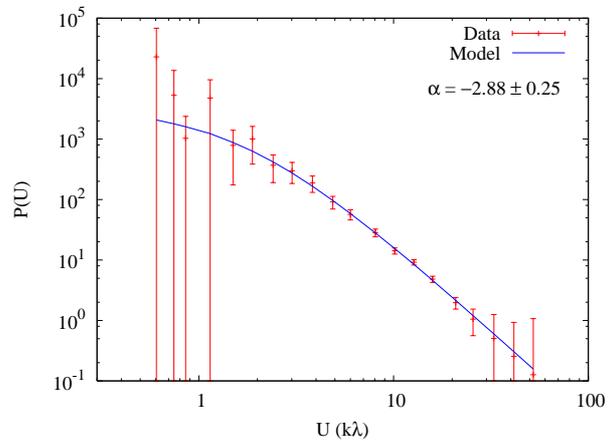}
\end{center}
\vspace{-0.3cm}
\caption{\label{fig:lineps} 
\small{Estimated intensity fluctuations power spectrum for one spectral channel using visibility data from the GMRT observation of H~{\sc i} absorption towards Cas~A. The solid line is the best-fit model with a power law power spectrum for the opacity fluctuations.}}
\end{wrapfigure} 
Here, angular brackets denote ensemble average over possible realization. If 
one assumes statistical isotropy, then $P(u,v)$ will be a function of only the 
magnitude $U = \sqrt{u^2+v^2}$. It is important to note that, due to the scale 
free nature of the turbulence, the power spectrum is expected to have a power 
law form $P(U) = A\,U^{\alpha}$ over the inertial range of scales. The 
structure function then will also be a power law \citep{lee75}, with power law 
index $\beta = -(\alpha+2)$.

One advantage of interferometric observations is that the observed 
visibilities are already measurements of Fourier transform components of the 
sky brightness. So, one can estimate the power spectrum from the observed 
visibilities without going into the complications and limitations of the 
imaging and deconvolution algorithms \citep{crov83,gree93,bhar01}. Based on 
this principle, a suitable method has been developed for radio interferometers 
like the GMRT \citep{nroy09}, and is used to estimate, directly from the 
visibility data, the H~{\sc i} opacity fluctuations power spectrum towards the 
extended background emission from the supernova remnant Cassiopeia~A (Cas~A). 
Figure~\ref{fig:lineps} shows the derived power spectrum for one of the 
velocity channels, and the best fit model overlaid on the data. It was 
established that the H~{\sc i} opacity fluctuations have a power law power 
spectrum with $\alpha = -2.86 \pm 0.10 (3\sigma)$ over the scale of $\sim 500$ 
AU to $2.5$ pc \citep{nroy10}. The line of sight towards Cas~A probes deep 
absorption from the local arm and the Perseus arm of the Milky Way. There was 
no significant difference in $\alpha$ for these two spiral arms. The results 
are in good agreement with other studies suggesting $\alpha \sim -2.7$ to $-3$ 
for the diffuse neutral medium \citep{crov83,desh00,dutt09}. Although the power 
law behaviour suggests a Kolmogorov-like turbulence, the power law index is 
significantly shallower than the Kolmogorov $-11/3$ spectrum for incompressible 
turbulence \citep{kolm41}. This mismatch may be due to the fact that even small 
ionization fraction can couple the gas with the interstellar magnetic field. In 
that case, the turbulence will be magnetohydrodynamic (MHD) in nature, and the 
density fluctuation power spectrum may have a different power law index (see 
\citealt{nroy09}, and references therein for a detailed discussion).

\begin{wrapfigure}{r}{0.56\textwidth}
\vspace{-0.8cm}
\begin{center}
\includegraphics[angle=-90, width=0.55\textwidth]{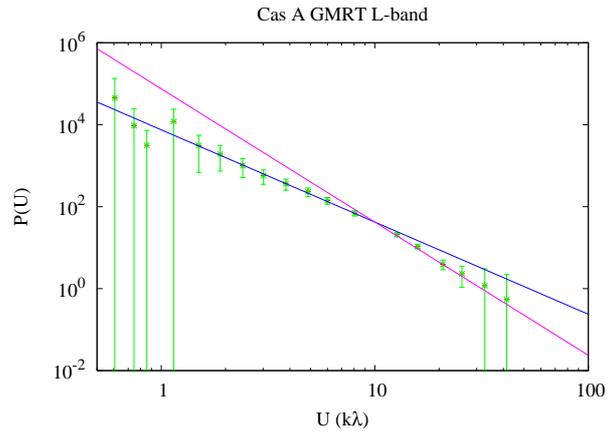}
\end{center}
\vspace{-0.3cm}
\caption{\label{fig:contps} 
\small{Power spectrum of the synchrotron emission intensity fluctuations for Cas~A at 1.4 GHz estimated from the GMRT observation. The power law index changes from $-2.22$ to $-3.23$ at $U\gtrsim10~k\lambda$.}}
\end{wrapfigure}

Apart from the density fluctuations, these data were also used to constrain 
the velocity fluctuations using ``velocity channel analysis'' \citep{laza00}. 
However, the effect of the velocity width of the spectral channels was found 
to be insignificant on the power law index. This result only put a weak 
constrain on the velocity structure function index $\beta = 0.2 \pm 0.6$, 
broadly consistent with \citet{kolm41} prediction of $\beta = 2/3$.

It is worth mentioning here that this study also revealed the presence of MHD 
turbulence in the supernova remnant from the continuum image of the background 
source Cas~A. This is revealed by the power spectra of the synchrotron 
radiation intensity fluctuations derived from the GMRT 1.4 GHz observation of 
Cas~A (see Figure~\ref{fig:contps}), and also from the VLA 5 GHz archival data 
for Cas~A and Crab Nebula. For Crab Nebula, the power law index is $\approx 
3.23$ over the whole range of $U$ probed in these studies. For Cas~A, the 
power spectrum has the same slope at large $U$. However, below $\sim 10~ 
k\lambda$, there is a break in the spectrum, and the index changes to $-2.22$. 
This break and flattening below $10~k\lambda$ (25 arcsec; an angular scale 
that corresponds to the shell thickness of Cas A) is interpreted as a 3-D to 
2-D transition of the turbulence in the supernova shell \citep{nroy09}.

Another way to get a handle on the nature of turbulence is to study the scaling 
of turbulent velocity dispersion with size of clouds. Similar studies for the 
molecular clouds were earlier reported by \citet{lars81} and others. As 
mentioned before, for the dense CNM phase, $T_s$ is a good proxy for $T_k$ (see 
sub-sections titled ``H~{\sc i} 21 cm radiation'' and ``Derived physical 
quantities'' for details). So, from the H~{\sc i} emission-absorption studies, 
one can estimate the non-thermal velocity dispersion $v^2_{\rm NT} \propto 
(T_{k,\rm max} - T_k) \approx (T_{k,\rm max} - T_s)$. Most of the time, it is 
not straightforward to estimate the distance of the cloud, and hence the size, 
directly. But, if the thermal pressure $P = nkT_k$ is assumed to be more or 
less constant, then the typical length scale of a cloud
\begin{equation}
L \sim N({\rm H~I})/n \sim  N({\rm H~I})/(P/kT_k) \propto N({\rm H~I}) T_s \,.
\end{equation} 

\begin{wrapfigure}{r}{0.56\textwidth}
\vspace{-0.8cm}
\begin{center}
\includegraphics[angle=-90, width=0.55\textwidth]{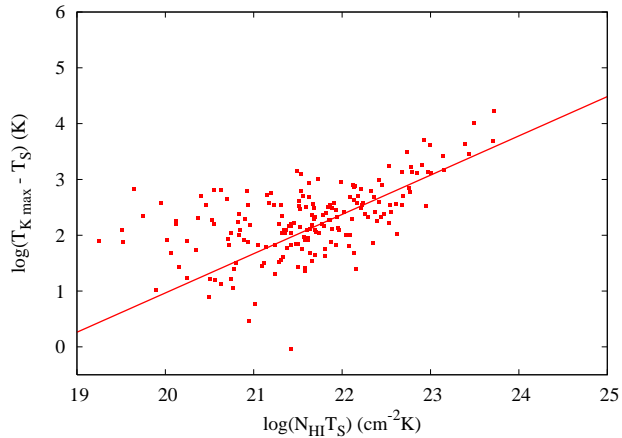}
\end{center}
\vspace{-0.3cm}
\caption{\label{fig:vntl} 
\small{Scaling of non-thermal line width with length scale derived from the millennium Arecibo 21-cm absorption-line survey results. A power law model with power index $2/3$ is shown as a solid line.}}
\end{wrapfigure}

This reasoning was used to derive the scaling of the form $v_{\rm NT} \sim 
L^\beta$ using the published millennium Arecibo 21-cm absorption-line survey 
results \citep{ht03a,ht03b}. The results are summarized in 
Figure~\ref{fig:vntl}. Over almost two orders of magnitude length scale, the 
data is well-represented by $\beta = 0.7 \pm 0.1$ \citep{nroy08}. For 
$N({\rm H~I}) T_s \lesssim 10^{21.5}{\rm ~cm}^{-2}{\rm K}$, the spread is 
higher, plausibly due to systematic effects of the spectral and spatial 
resolution of the Arecibo telescope. The result is consistent with the $2/3$ 
scaling predicted by \citet{kolm41}, who derived this relation by assuming 
that the dissipation rate is scale invariant over the inertial range. Note 
that this correlation, whether or not actually related to a Kolmogorov-like 
turbulence, can be used to estimate the non-thermal velocity dispersion and, 
in turn, the true kinetic temperature, taking it as a purely phenomenological 
model.\\


\noindent{\bf \Large Temperature of the diffuse neutral ISM}\vspace{0.25cm}

\noindent In the classical model of multiphase ISM, the balance of heating and 
cooling rate leads to the two stable ``phases'' (CNM and WNM) of the neutral 
ISM. Gas at the intermediate kinetic temperature ($T_k \sim 500 - 5000$ K) will 
suffer from runaway heating or cooling to evolve into either the WNM or the CNM 
respectively. So, in this model, H~{\sc i} at intermediate temperature exists 
``only as a transient phenomenon'', and does not contribute significantly to 
the total column density \citep{fiel69,mcke77,wolf03}. From an 
observational point of view, H~{\sc i} emission-absorption studies have shown 
the presence of narrow CNM components with high opacity, and with inferred 
$T_k$ in the expected range of $\lesssim 200$ K \citep{radh72,ht03a,nroy06}. 
The wide emission components with very little or no corresponding absorption 
are assumed to be originating from the WNM (see Figure~\ref{fig:spec} and the 
related discussion in the ``Derived physical quantities'' sub-section). Due to 
their low optical depth, measurements of the WNM spin or kinetic temperature 
are rare \citep{cari98,dwar02,heil01,begu10,murr14}. However, the thermal 
steady state scenario was considered to be the correct model broadly in 
agreement with the data.

Over the last few years, there were observational indications of the presence 
of a significant amount of H~{\sc i} with temperature in the unstable range. 
These results seriously challenged the status of the standard model. 
Unfortunately, only a few of these measurements are from interferometric 
observations \citep{kane03,brau05}, and the rest comes from single dish 
observations \citep[Millennium Arecibo 21-cm absorption-line survey;][]{ht03b}. 
The main issue with such single dish observations is that the effect of the 
H~{\sc i} emission must be modelled accurately to derive the absorption 
spectra. This, and the contamination of the spectra by unwanted emission from 
the main beam may have significant systematic effects, and make the results 
less reliable. The other potential issue is that, for WNM, neither $T_s$ nor 
$T_{k,\rm max}$ is a reliable proxy for $T_k$; the non-thermal broadening may 
be significant, and also $T_s$ may considerably differ from $T_k$ 
\citep{lisz01}. Finally, in presence of CNM components along the line of sight, 
``self-absorption'' of the background emission by the foreground components may 
make modelling the emission spectrum, if not impossible, really difficult. Even 
considering these possible observational shortcomings, the main result, that a 
large fraction of gas is in the ``unstable'' phase, has serious implications 
for the thermal steady state model.

\begin{wrapfigure}{r}{0.56\textwidth}
\vspace{-0.8cm}
\begin{center}
\includegraphics[angle=-90, width=0.55\textwidth]{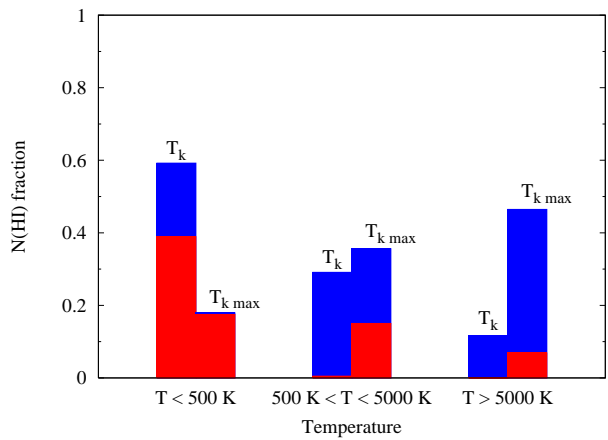}
\end{center}
\vspace{-0.3cm}
\caption{\label{fig:amsh} 
\small{Distribution of H~{\sc i} in different phases based on temperatures computed without and with correcting the non-thermal broadening (bars marked with $T_{k,\rm max}$ and $T_k$ respectively). Red (blue) are H~{\sc i} emission components that are detected (not detected) in absorption. Note that the correction implies only a small amount of gas in the stable WNM phase; however, the unstable fraction remains almost unchanged.}}
\end{wrapfigure}

It is important to note that the \citet{ht03b} draw the conclusion of a 
significant fraction of unstable gas based on $T_{k,\rm max}$ derived from the 
observed emission line widths. It is possible to use a Kolmogorov-like scaling 
(see Figure~\ref{fig:vntl}) to estimate the turbulent velocity dispersion, and 
derive the true temperature distribution. A reanalysis of the \citet{ht03b} 
data implementing this correction noticeably alters the distribution 
\citep{nroy09}. Figure~\ref{fig:amsh} shows the main result -- the distribution 
of $N({\rm H~I})$ at different temperature ranges before and after correcting 
the non-thermal broadening. As expected, a lot of ``warm'' gases based on 
$T_{k,\rm max}$ are indeed just cold gas with significant non-thermal 
broadening. There is, in fact, very little ($\approx 10\%$) H~{\sc i} in the 
stable warm phase in this sample. However, even after this correction, a 
similar fraction of gas ($\approx 30\%$) remains in the unstable phase 
\citep{nroy09}.

\begin{figure}[t]
\begin{center}
\includegraphics[width=5cm,height=7.5cm,angle=-90.0]{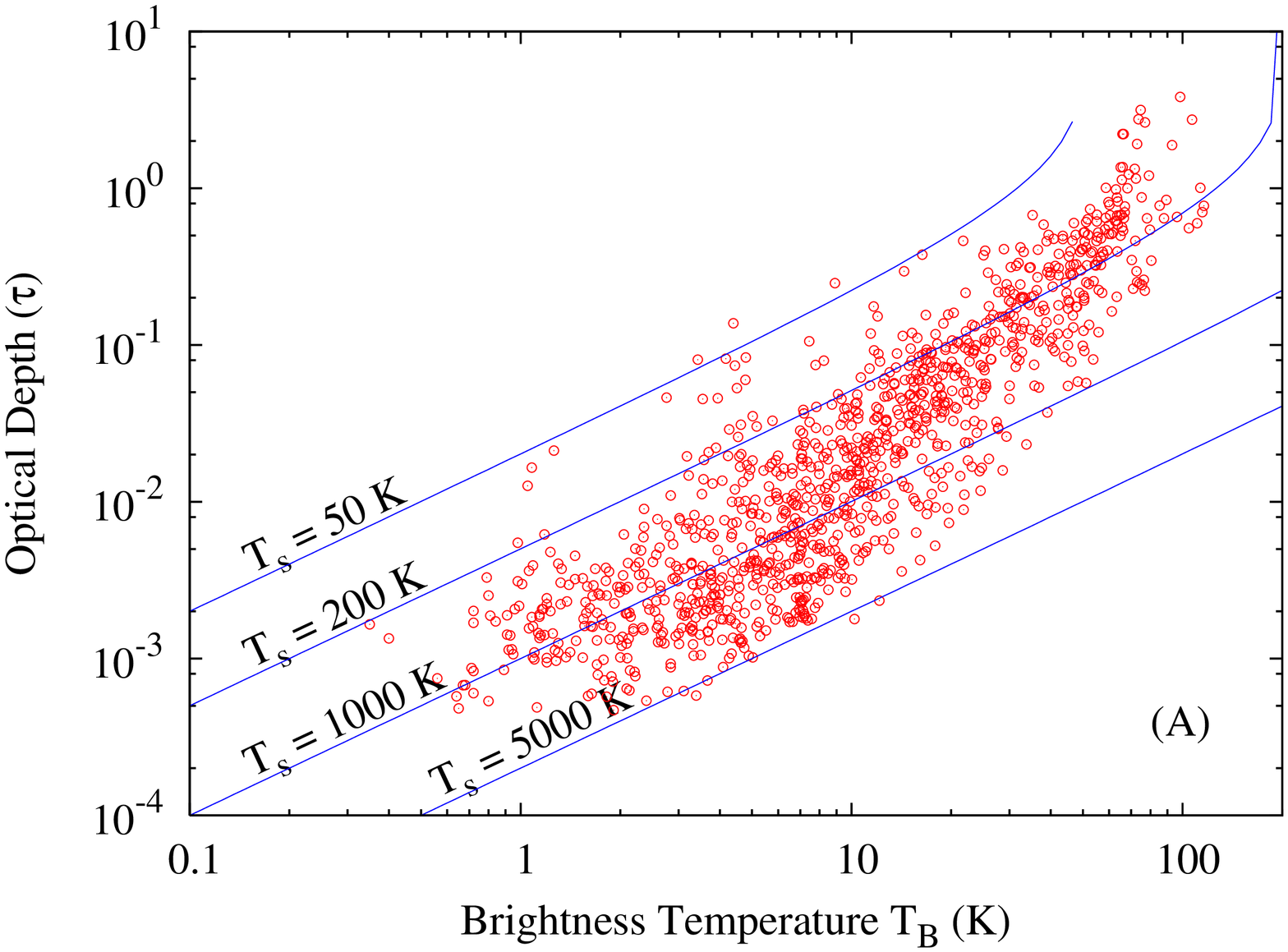}\includegraphics[width=5cm,height=7.5cm,angle=-90.0]{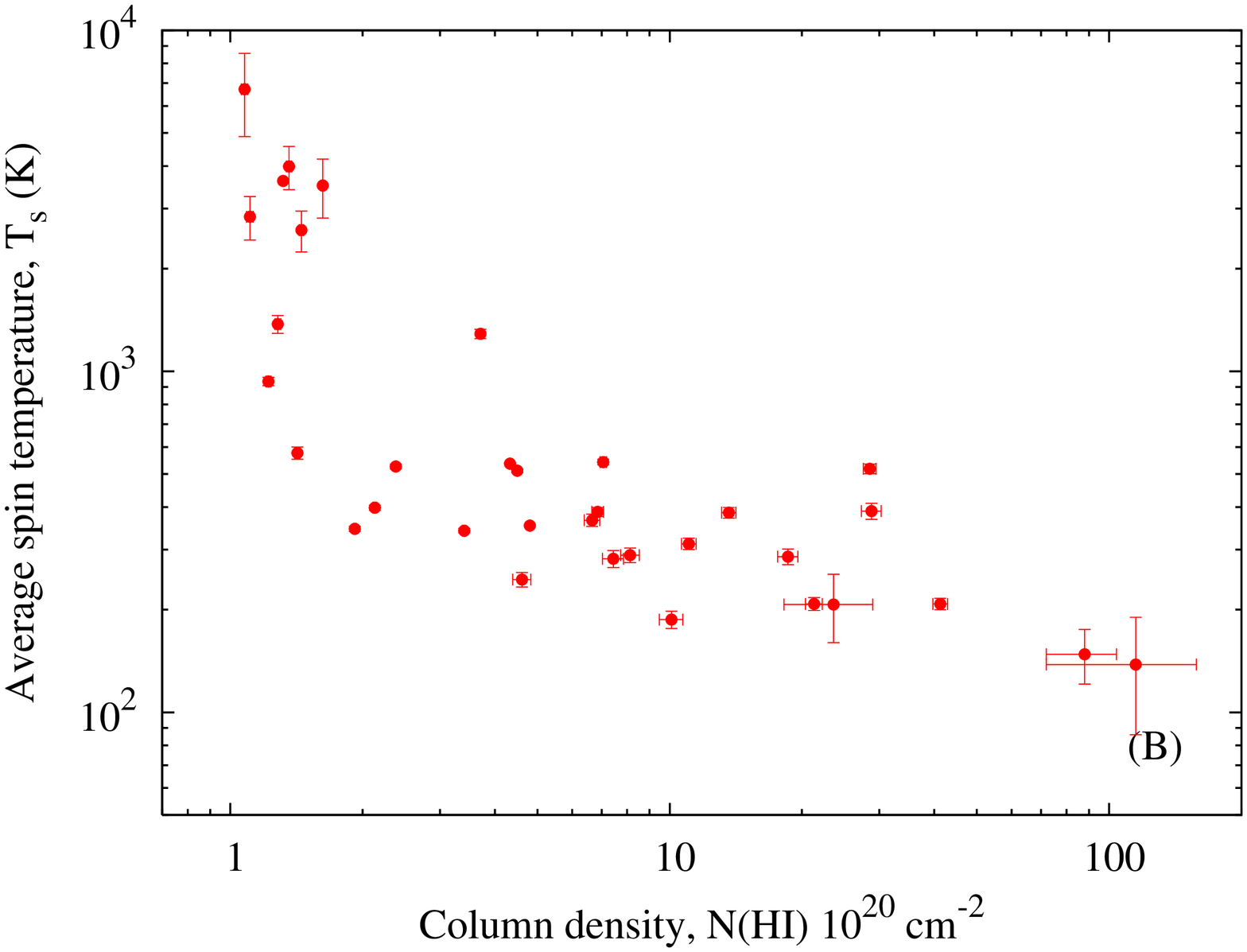}
\caption{\label{fig:nhts} 
\small{(A): H~{\sc i} 21 cm optical depth at $1$ km~s$^{-1}$ resolution from the absorption survey, and the corresponding $T_B$ from the LAB survey. Solid lines mark different $T_s$ values. Channels with intermediate $T_s$ may be a mix of cold and warm gas. (B): Integrated column density and average spin temperature for all 34 lines of sight. The high $T_s$ values imply presence of warm gas.}}
\end{center}
\vspace{-0.5cm}
\end{figure}

The fact that, due to various systematic effects, the single dish absorption 
spectra remains less reliable (particularly for weak and wide WNM components), 
prompted us to also start an interferometric survey with the GMRT and the WSRT. 
The main aim of this ongoing survey is to acquire high sensitivity, high 
spectral resolution Galactic H~{\sc i} 21 cm absorption towards a large number 
of background sources. A large sample, and a high sensitivity (RMS opacity 
$\sim 10^{-3}$) are necessary to draw a statistically significant conclusion 
about the WNM properties. So far, $\sim 50$ lines of sight have been observed 
($6 - 8$ hr time per source). Details of the survey, the analysis procedure 
and results based on a brighter subsample of $34$ sources, including two 
southern sources observed with the Australian Telescope Compact Array (ATCA), 
have been reported recently \citep{nroy13a,nroy13b}. One example absorption 
spectrum from this survey is shown in Figure~\ref{fig:spec}.

From each of these lines of sight, using absorption spectra from this survey 
and the corresponding emission spectra from the LAB survey, $T_s$ values are 
computed at $\sim 1$ km~s$^{-1}$ velocity resolution for each channel with 
$\geq 3\sigma$ detection of H~{\sc i} absorption. As shown in Figure 
\ref{fig:nhts} (A), many channels have $T_s > 200$ K, clearly an indication of 
a mix of cold and warm gas. The line of sight average $T_s$, plotted in 
Figure~\ref{fig:nhts} (B), also have values $> 200$ K for most of the cases. 
Based on $T_s$ at $1$ km~s$^{-1}$ resolution, about $\sim 50\%$ of the gas (by 
column density) has $T_s < 200$ K (i.e. channels with large cold gas fraction), 
and $> 10\%$ has $T_s < 1000$ K (possibly with large warm gas fraction). The 
rest, $\sim 40\%$ of the gas, with $200 {\rm ~K} < T_s < 1000 {\rm ~K}$ is not 
necessarily ``unstable'' gas, but can be only a mix of CNM and WNM. However, 
even without Gaussian decomposition of the profiles, based on the average $T_s$ 
values and the sensitivity limit of the survey, one can conclude that the 
detected H~{\sc i} absorption is not entirely from cold gas with $T_k < 200$ K, 
and must also have contribution from (stable or unstable) warm gas 
\citep{kane11,nroy13a,nroy13b}. An upper limit of the CNM fraction was derived 
assuming that all cold gas has $T_s = 200$ K. The median CNM fraction was found 
to be $f_{\rm CNM} \approx 0.52$ for the sample.

Finally, Gaussian decomposition of the absorption profiles were carried out, 
using the Levenberg-Marquardt $\chi^2$-minimization, to estimate $T_{k,\rm 
max}$ of the individual clouds along the lines of sight. It was found that the 
number of components with $T_{k,\rm max} > 5000$ K is very few (only seven in 
this range, and, considering the measurement uncertainty, possibly a few more 
out of a total 214 components). As the survey has adequate sensitivity to 
detect H~{\sc i} in absorption even at this temperature range, one can put a 
strong constraint on the fraction of gas in the ``classical'' warm phase. 
Combining the fact (i) that only a few components have $T_{k,\rm max}$ in the 
range for stable WNM, (ii) that a significant fraction of gas can not be below 
the detection limit of the survey, and (iii) that the median $f_{\rm CNM}$ is 
at most about $0.5$, the lower limit of unstable gas fraction was found to be 
$\sim 0.3$. We conclude that, indeed, a significant fraction ($\gtrsim 30\%$) 
of the gas is in the so-called unstable phase.\\


\noindent{\bf \Large Discussions and Conclusions}\vspace{0.25cm}

\noindent Turbulence is an important ingredient of ISM physics in various 
regards (see \citealt{elme04}, and \citealt{scal04}, for comprehensive reviews 
on observations and implications of the ISM turbulence). The small scale 
structures of the ISM, originating from turbulence, seems to be ubiquitous not 
only in the Milky Way, but also in other galaxies 
\citep{west99,elme01,dutt09,sria13}. From the galaxy disk scale to molecular 
cloud, it affects the ISM dynamics and properties \citep{rome10,shad11,hoff12}. 
The results reported in this article, based on H~{\sc i} observations, show 
that both the density and velocity fluctuation statistics are consistent with a 
near-Kolmogorov turbulence in the diffuse ISM of our Galaxy. A working model is 
also proposed to estimate the non-thermal broadening of the line width and to 
derive the true kinetic temperature of the gas.

Apart from the H~{\sc i} 21 cm line, various other tracers at different 
wavelengths are also used to study turbulence in different phases of the ISM 
\citep{welt94,gibs07,bois13}. However, neutral hydrogen, which is present 
almost everywhere in the Galaxy, can be used to probe turbulence over a wide 
range of scales, using multiple observational techniques (from single dish 
emission study to VLBI small scale absorption measurement). For example, VLBI 
observations can constrain the density and the velocity power spectra at 
milliarcsec scales \citep{nroy12,dutt14}. Constraining the turbulent behaviour 
of the ISM at the scale of VLBI observation can, in near future, address 
whether the same power law behaviour of the energy cascade is observed at pc 
and AU scales, and also constrain the energy dissipation scale and mechanism. 
It will also be useful to see, how the VLBI observations at AU scale compare 
with observations of tiny scale structures detected using multiwavelength 
observations of other tracers .

Recent Galactic H~{\sc i} observations are also questioning the classical 
thermal steady state model of the multiphase ISM. There is clear evidence that 
$\gtrsim 30\%$ of the diffuse neutral gas is in the unstable phase. However, 
any effort of understanding the properties of the multiphase ISM should also 
take into account the possible effects of the turbulence. Observationally, one 
must get a handle on the turbulent velocity dispersion to accurately estimate 
the temperature, and to critically examine the validity of thermal steady 
state model. From theoretical consideration, on the other hand, turbulence may 
potentially play a role in creating and supporting gas at the intermediate 
temperature range. Recent numerical studies suggest that turbulence can explain 
the presence of some amount of gas in the unstable warm phase 
\citep{audi05,audi10}. These numerical simulations, however, still predict a 
large fraction of gas to be in the stable WNM phase, somewhat in contrast to 
the observational results. More recent hydrodynamic simulations by 
\citet{kim14} can explain some of the observational results of 
\citet{nroy13a,nroy13b}. It remains interesting to see whether more data and 
further numerical studies will result in a better agreement between the details 
of the observational results and the theoretical predictions.\\

\newpage


\noindent{\bf \Large Acknowledgements}\vspace{0.25cm}

\noindent The author thanks the Indian National Science Academy (INSA), and 
Prof. Lakhotia, Editor-in-Chief, for giving him a chance of writing this 
article for the Proc. of the INSA. The author is grateful to Aritra Basu for 
useful comments on an earlier version of the paper.

The author acknowledges his collaborators Somnath Bharadwaj, Robert Braun, 
Jayaram N. Chengalur, Prasun Dutta, Leshma Peedikakkandy, and Nissim Kanekar 
for their contribution, help and support in various projects which form the 
basis of this review article. The author acknowledges support from the 
Alexander von Humboldt Foundation, and also acknowledges support from the 
National Centre for Radio Astrophysics (NCRA) of the Tata Institute of 
Fundamental Research (TIFR) during his Ph.D., when a significant fraction of 
this work was done. Photograph of the GMRT antennas is provided by Arup Biswas, 
and those of the Effelsberg telescope and the LOFAR station are from Susmita 
Chakravorty.

The results reported here are based on data mainly from observations with the 
GMRT and the WSRT, and analysis carried out using the NRAO AIPS. The author 
thanks the staff of the GMRT and the WSRT who have made these observations 
possible. The GMRT is run by NCRA-TIFR. The WSRT is operated by ASTRON (the 
Netherlands Institute for Radio Astronomy), with support from the Netherlands 
Foundation for Scientific Research (NWO). The NRAO is a facility of the 
National Science Foundation (NSF) operated under cooperative agreement by 
Associated Universities, Inc. (AUI). Some results are also derived from the 
ATCA observations, the VLA archival data, the Leiden/Argentine/Bonn Galactic 
H~{\sc i} Survey data, and the millennium Arecibo 21-cm absorption-line survey 
data. This research has also made use of the NASA's Astrophysics Data System.

\end{document}